\documentclass[letterpaper,twoside,english,aps,prl,superscriptaddress]{revtex4}
\usepackage{ae,aecompl}

\usepackage[T1]{fontenc}
\usepackage[latin9]{inputenc}
\usepackage{amsmath}
\usepackage{amssymb}
\usepackage{graphicx}
\usepackage{esint}

\makeatletter

\newcommand{\noun}[1]{\textsc{#1}}

\@ifundefined{textcolor}{}
{%
 \definecolor{BLACK}{gray}{0}
 \definecolor{WHITE}{gray}{1}
 \definecolor{RED}{rgb}{1,0,0}
 \definecolor{GREEN}{rgb}{0,1,0}
 \definecolor{BLUE}{rgb}{0,0,1}
 \definecolor{CYAN}{cmyk}{1,0,0,0}
 \definecolor{MAGENTA}{cmyk}{0,1,0,0}
 \definecolor{YELLOW}{cmyk}{0,0,1,0}
 }

\@ifundefined{definecolor}{\usepackage{color}}{}
\newcommand{\dagga}{{\phantom{\dagger}}}

\usepackage{babel}

\makeatother

\usepackage{babel}
\begin{document}

\title{Localization and Glassy Dynamics Of\\
 Many-Body Quantum Systems }

\author{Giuseppe Carleo({*})}

\affiliation{\noun{SISSA} -- Scuola Internazionale Superiore di Studi
Avanzati \\
 and CNR-IOM \noun{DEMOCRITOS} Simulation Center, Via Bonomea 265
I-34136 Trieste, Italy}

\author{Federico Becca}

\affiliation{\noun{SISSA} -- Scuola Internazionale Superiore di Studi
Avanzati \\
 and CNR-IOM \noun{DEMOCRITOS} Simulation Center, Via Bonomea 265
I-34136 Trieste, Italy}

\author{Marco Schir\'{o}}

\affiliation{Princeton Center for Theoretical Science and Department of Physics,
Princeton University, Princeton, New Jersey 08544, USA }

\author{Michele Fabrizio}

\affiliation{\noun{SISSA} -- Scuola Internazionale Superiore di Studi
Avanzati \\
 and CNR-IOM \noun{DEMOCRITOS} Simulation Center, Via Bonomea 265
I-34136 Trieste, Italy}
\affiliation{\noun{ICTP} -- The Abdus Salam International Center 
for Theoretical Physics, P.O. Box 586, I-34151 Trieste, Italy}
 
\begin{abstract}
When classical systems fail to explore their entire configurational
space, intriguing macroscopic phenomena like aging and glass formation
may emerge. Also closed quanto-mechanical systems may stop wandering
freely around the whole Hilbert space, even if they are initially
prepared into a macroscopically large combination of eigenstates.
Here, we report numerical evidences that the dynamics of strongly
interacting lattice bosons driven sufficiently far from equilibrium
can be trapped into extremely long-lived inhomogeneous metastable
states. The slowing down of incoherent density excitations above a
threshold energy, much reminiscent of a dynamical arrest on the verge
of a glass transition, is identified as the key feature of this phenomenon.
We argue that the resulting long-lived inhomogeneities are responsible
for the lack of thermalization observed in large systems. Such a rich
phenomenology could be experimentally uncovered upon probing the 
out-of-equilibrium dynamics of conveniently prepared quantum states of trapped
cold atoms which we hereby suggest.
\end{abstract}
\maketitle
The ergodicity axiom in classical statistical mechanics states that,
during its time evolution, a closed macroscopic system uniformly explores
the entire phase space compatible with conservation laws, so that
the time average of any observable comes to coincide with the micro-canonical
ensemble average and, when the observable is local, also with the
canonical Gibbs ensemble average. Nonetheless, ergodicity can be violated
in classical systems, a noticeable example being 
glasses.~\cite{ParisiMezard1987}
Quantum effects might also spoil ergodicity by preventing the wave
function from diffusing within all available configurations. This
phenomenon is actually known to occur in the presence of disorder
and manifests itself either by single-particle~\cite{Anderson1958}
or many-particle~\cite{Basko2006,Oganesyan2007,Pal2010} wave function
localization. However, alike classical models for glassy behavior,
ergodicity breakdown in the quantum dynamics may not necessarily require
disorder and it could instead be entirely due to frustrating dynamical
constraints.~\cite{Biroli2001,Foini2011} This issue is currently
attracting great interest,~\cite{Olshanii_Rigol_nature,SilvaRMP}
since well controlled realizations of closed quantum systems have
become feasible upon trapping cold atomic species.~\cite{Bloch2008}
Indeed, similarly to what can be done in numerical simulations, one
can prepare atoms in a given initial state and probe their time evolution
under a Hamiltonian whose parameters are fully under control, thus
offering the unique opportunity to monitor the ergodicity principle
at work in the quantum realm.~\cite{VonNeumann_1929}

In this work, we report numerical evidences that an isolated system
of strongly interacting bosons, modeling atoms in optical lattices,
can be trapped during its evolution into long-lived inhomogeneous metastable 
states, provided that its internal energy exceeds a certain threshold. 
We argue that the slowing down of high-energy incoherent excitations in 
the strongly correlated system is the key feature responsible
for this dynamical arrest, much resembling a kind of glass transition.
By formulating the problem in a different language, we explicitly
show that a system initially prepared in a inhomogeneous state is
unable to diffuse within the entire configurational space; such a
dynamical localization in the many-body Hilbert space looks intriguing
and may represent a kind of many-body Anderson localization~\cite{Basko2006}
that occurs without disorder. The above phenomenon is put in further
relation with and deemed responsible for the lack of ergodicity observed
in large finite size systems. This belief is confirmed by means of
a novel time-dependent variational Monte Carlo method that we introduce
hereby and an experimental set up to uncover this very rich phenomenology
is also suggested.

One of the simplest models that can be realized in experiments is
the Bose-Hubbard Hamiltonian:~\cite{Bloch2008,Zoller1998} 
\begin{equation}
\mathcal{H}=-J\sum_{\left\langle i,j\right\rangle }\left(b_{i}^{\dagger}b_{j}^{\dagga}+\text{h.c.}\right)+\frac{U}{2}\sum_{i}\, n_{i}(n_{i}-1),\label{eq:Bosehubb}
\end{equation}
 characterized by the amplitude $J$ for a bosonic species of an atom
to hop between nearest-neighboring wells of an optical lattice and
by a \textit{local} repulsion $U$ among atoms localized in the same
potential well. The operators $b_{i}^{\dagger}$ and $b_{i}^{\dagga}$
create and destroy, respectively, a boson on site $i$, and $n_{i}=b_{i}^{\dagger}b_{i}^{\dagga}$
is the density operator.~%
\footnote{In the experimental setup, an additional confining potential is 
usually present. We do not expect the latter to play a major role in what
we shall discuss.%
} Experiments are often performed with anisotropic lattices that realize
a collection of almost uncoupled chains, a fortunate case for numerical
simulations that we shall mainly consider hereafter, 
apart from a brief excursion in two dimensions towards the end of the paper. 

In one dimension, obstacles to ergodicity can arise in integrable
models.\cite{Calabrese2007} However, the Hamiltonian \eqref{eq:Bosehubb} 
is not integrable and, indeed, there are experimental evidences 
that its dynamical evolution may succeed in fast 
relaxing to a {\sl thermal} state.  
Specifically, in a recent experiment a system of $^{87}$Rb atoms,
well described by the Hamiltonian~\eqref{eq:Bosehubb}, has been prepared
in a state in which the sites of the optical lattice were alternatively
empty and singly occupied.~\cite{Bloch2011} This state was let evolve
for different experimental conditions corresponding to 
different ratios $U/J$.
Even at the largest value $U/J\simeq10$, the initial density profile
$(\dots1,0,1,0,\dots)$ was found to rapidly relax to the 
homogeneous \textit{thermal}
one $(\dots\frac{1}{2},\frac{1}{2},\frac{1}{2},\frac{1}{2},\dots)$, with half 
a boson per site, 
much faster than the integrable counterparts of non-interacting or
infinitely-interacting (i.e., hard-core) bosons and consistently with
the increased number of relaxation channels that opens once integrability
is lost. 

Numerical simulations of the above experiment successfully reproduce 
the observed \textit{thermal} behavior.~\cite{Bloch2011} 
However, there are also 
numerical evidences pointing out a breakdown of ergodicity in 
the same model \eqref{eq:Bosehubb} but within a different region of the 
parameter 
space, specifically when the number of bosons is one per site. 
This case, as well as any other one with integer density, is special because, 
at equilibrium and at zero temperature, the model \eqref{eq:Bosehubb} undergoes 
a quantum phase transition into a Mott insulator above a critical $U/J$. 
Even though the phase transition is washed out by thermal fluctuations, 
nonetheless its influence on the spectrum seems to prevent thermalization 
above a certain $U/J$ in numerical simulations of finite size 
systems\cite{Kollath2007,Roux2009}; a result that may~\cite{Cassidy2009} 
or may not~\cite{Kollath_Biroli2010} prelude to a true breakdown of 
ergodicity in the actual thermodynamic limit.  
 
Here, we propose an experiment a lot alike the one previously 
described,~\cite{Bloch2011} which may distinguish 
very sharply a change from chaotic to non-chaotic 
dynamics in the Bose-Hubbard model \eqref{eq:Bosehubb}.
Our proposal starts by observing that, in the gapless phase 
next to the Mott insulator at one boson per site, 
low-energy itinerant Bogoliubov quasiparticles 
should coexist with high-energy incoherent excitations, which, for $U/J\gg1$, 
can be identified as sites occupied by more than a single boson.
It is well possible that the overall relaxation depends critically upon 
the population of those high-energy excitations,~\cite{Rosch2008} 
which can be assessed by tailoring initial states with lots    
of doubly occupied sites rather than none as in 
the experiment of Ref.~\cite{Bloch2011}. This is indeed what we 
propose and simulate by means of different and complementary numerical tools,
such as exact diagonalization, time-evolving block decimation~\cite{Vidal2004}
and a novel time-dependent variational Monte Carlo algorithm, as discussed
in more detail in the Supplementary Material.

\subsection*{Results}

\subsubsection*{Inhomogeneous initial states and dynamical localization }

\begin{figure}[t]
 \includegraphics[width=0.48\paperwidth]{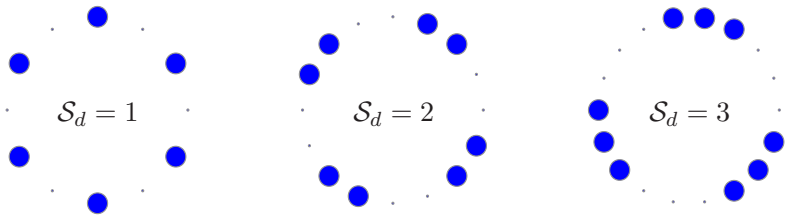} 
\caption{\label{fig:DefType} Inhomogeneous initial states constituted by 
clusters of doubly occupied sites (large blue circles) and empty sites (small
dots). The size of each cluster is denoted by $\mathcal{S}_{d}$.}
\end{figure}

We start from analyzing the model at density $n=1$, when at equilibrium
a Mott transition occurs at a critical $(U/J)_{c}\simeq3.5$.~\cite{Monien1998}
We imagine to prepare an initial state where all the sites are either
empty or doubly occupied. In particular we consider the states depicted
in Fig.~\ref{fig:DefType}, namely with clusters of doubly occupied
sites of variable size $\mathcal{S}_{d}$. These states are let evolve
with the spatially homogeneous Hamiltonian~\eqref{eq:Bosehubb} for
different $U/J$, below and above the critical value. While for small
$U/J$ the density profile rapidly reaches the equilibrium configuration
$(\dots1,1,1,1,\dots)$, for large $U/J$, it stays close to its initial
value for a remarkably long time. Eventually, since the system is
finite, the density profile approaches the homogeneous plateau, with
small residual oscillations that get damped as the system size increases.

\begin{figure*}
\includegraphics[width=0.4\paperwidth]{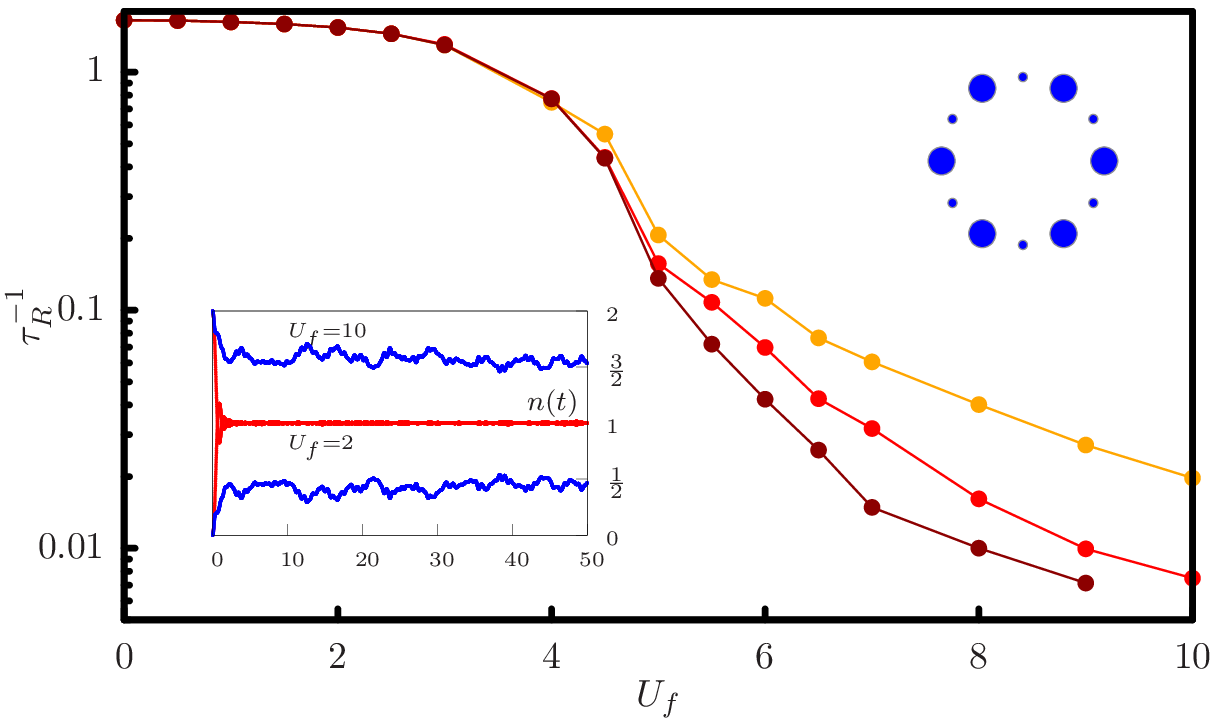}\hfill{}\includegraphics[width=0.4\paperwidth]{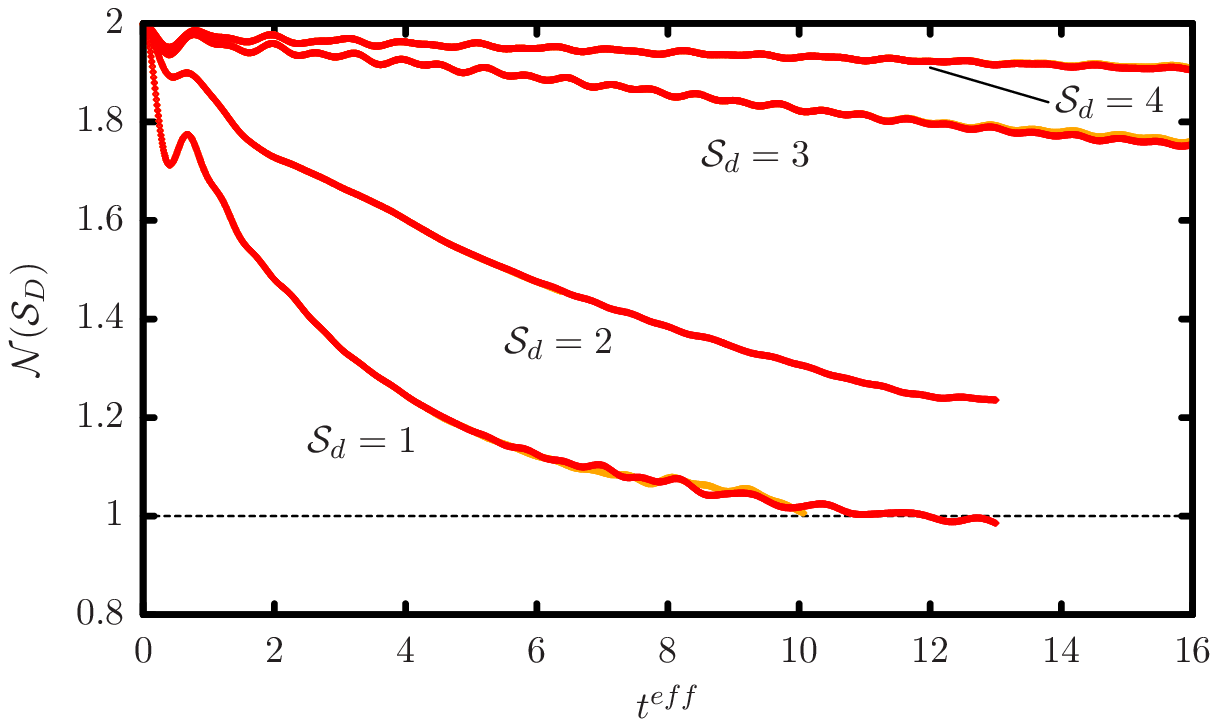}
\caption{\label{fig:DecDef} 
\emph{Left panel} -- Inverse relaxation times $\tau_{R}^{-1}$ of the local 
density for the initial state $(\dots2,0,2,0,\dots)$. Exact diagonalization 
results are reported, with darker points marking larger systems, respectively 
$N=8$, $10$ and $12$ with periodic-boundary conditions. In the Inset, the time dependence of the on-site densities is shown. 
\emph{Right panel} --
Effective-Hamiltonian evolution of the average density within a cluster of 
different size $\mathcal{S}_{d}$, where the dimensionless time is defined as 
$t^{eff}=\frac{2J^{2}}{U}t$. We show results for $N=72$ and $96$ sites with
open-boundary conditions, obtained by the time-evolving block decimation 
technique.~\cite{TebdOpen} We note that even small clusters of doubly
occupied sites can effectively freeze the dynamical evolution in the
large-$U$ regime.}
\end{figure*}

We can define a relaxation time $\tau_{R}$ (whose inverse is shown
in Fig.~\ref{fig:DecDef}, for $\mathcal{S}_{d}=1$), as the first
time for which the local density approaches its homogeneous value.
We highlight that $\tau_{R}$, at a specific $(U/J)_{c}^{dyn}$, has
a sudden step up, which becomes sharper and sharper as the system
size increases. The above results show that above $(U/J)_{c}^{dyn}$
the system has the tendency to stay dynamically trapped into long-lived
inhomogeneous configurations.

We emphasize that such a behavior is all the more remarkable not  
because doubly occupied sites seem unable to decay, which is known to 
require for $U/J\gg 1$ very long times and large system 
sizes,~\cite{Kollath_Biroli_Roux2010} but rather because they are not 
capable to move, hence restore translational symmetry.
We can better understand this surprising result 
by an effective Hamiltonian that can be derived
following the same reasoning of Refs.~\cite{Petrosyan2007,Rosch2008}.
For a sufficiently large interaction, it is justified to project the
evolution onto states with the same potential energy per site $U/2$,
at least for time scales shorter than $U^{2}/J^{3}.$ One realizes
that states with the same potential energy but with triply and singly
occupied sites start to contribute only at order $J^{4}/U^{3}$, so
that, with accuracy $J^{2}/U$, the number of doubly occupied sites
is conserved. If we associate a fictitious spin up or down to a doublon
(doubly occupied site) or a holon (empty site), respectively, we find
that the effective Hamiltonian that controls the evolution reads:
\begin{equation}
\mathcal{H}^{eff}=\frac{2J^{2}}{U}\,\sum_{\langle ij\rangle}\,\bigg[-8\, S_{i}^{z}\, S_{j}^{z}+\Big(S_{i}^{+}\, S_{j}^{-}+S_{i}^{-}\, S_{j}^{+}\Big)\bigg],\label{eq:Heff}
\end{equation}
 which describes a hard-axis 
ferromagnetic Heisenberg model.~\cite{Petrosyan2007} 
We note that the evolution of an XXZ spin chain starting from an antiferromagnetic ordered initial state has been studied numerically in Ref.~\cite{Barmettler2010} with DMRG. 
Here we consider the effective dynamics~(\ref{eq:Heff}) in a regime when the anisotropy would favor a ferromagnetic ground state and when the initial state contains clusters of up and down spins of increasing lenght. 
The smaller cluster size of  $\mathcal{S}_d=1$ corresponds to a Neel initial state considered in Ref~\cite{Barmettler2010}.
In the right panel of Fig.~\ref{fig:DecDef} we show the time-dependence
of the clusters density in the large $U$ regime, evolved according
to the effective Hamiltonian~\eqref{eq:Heff}. Remarkably, we see that
even for small clusters the system fails to restore the
spatial homogeneity up to very long time scales, which turn to be
far beyond those currently accessible in typical experimental setups.

The slowing down of the dynamics can be traced back to the 
effective attraction among doublons,~\cite{Rosch2008}
which makes their aggregates hard to break up. In other words, 
what seems to matter more is the dissociation of clusters of doublons, 
rather than the decay of a single one. Indeed we have checked (not shown)
that in the large $U$ regime the system has the tendency to get dynamically
stuck into clusters of doublons of finite size, whereas a much faster
annihilation and recombination rate is observed in the small $U$
limit.

\begin{figure*}
 \includegraphics[width=0.58\paperwidth]{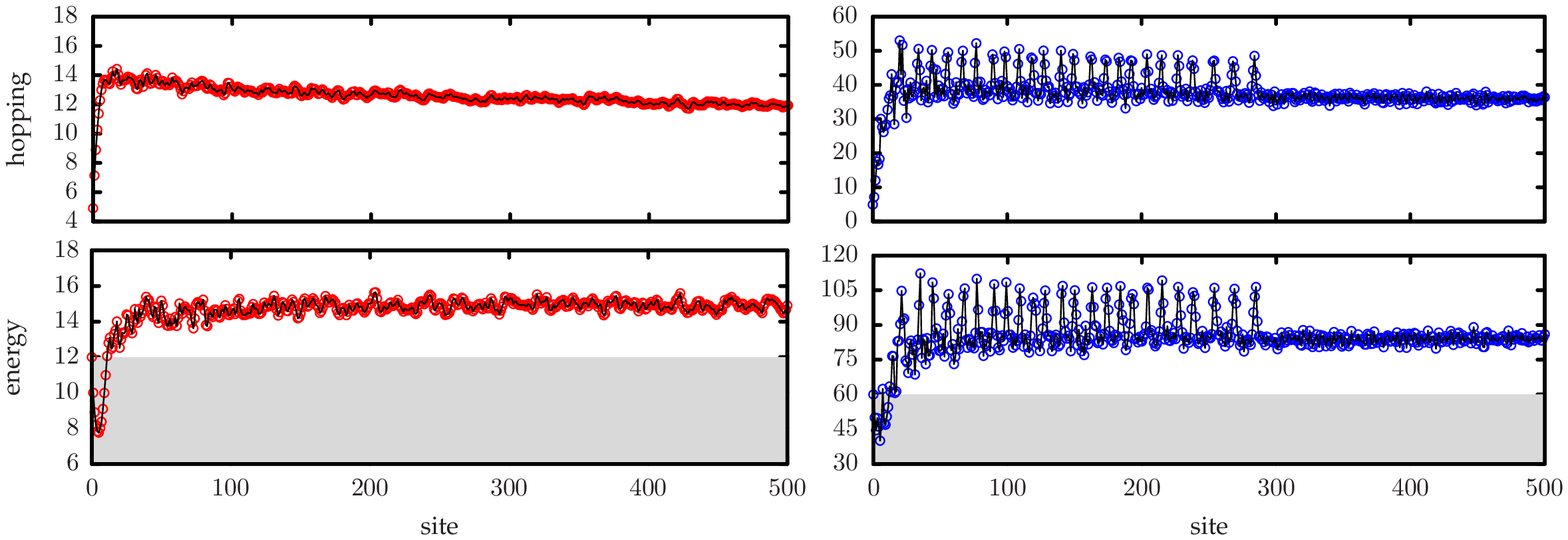}\hfill{}\includegraphics[width=0.22\paperwidth]{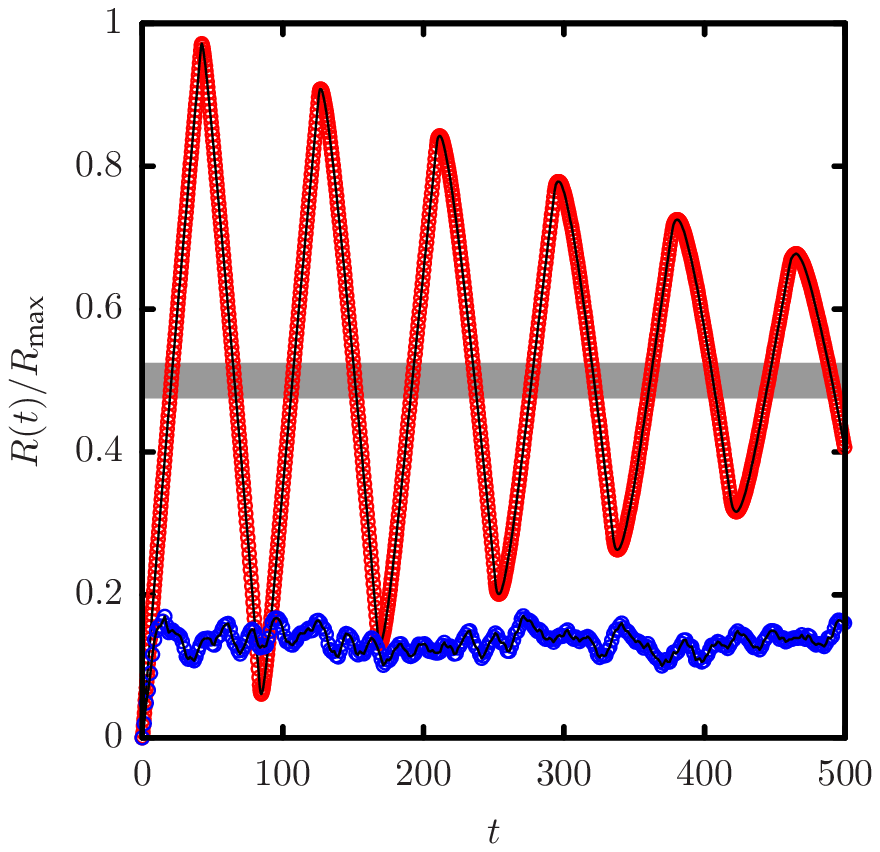}
\caption{\label{fig:1d} 
\emph{Left panels} -- On site energies and nearest-neighbor
hoppings of the effective chain that represents the Hamiltonian in
the Lanczos basis starting from the state $(\dots,2,0,2,0,\dots)$. 
Red points refer to $U=2J$, when the particle does diffuse
starting from site 0, while blue points to $U=10J$, when it does
not. The shaded regions correspond to energies less or equal that
of the initial state. \emph{Right panel} -- Time-dependent expectation
value of the wave-packet position of the effective particle traveling
in the Hilbert space generated by a chain of $R_{\text{max}}=1000$
Lanczos states. The red points correspond to $U=2J$ and the blue
points to $U=10J$ (the original lattice size of the Bose-Hubbard
model is $N=12$). The shaded region marks the center of the Lanczos
chain, which is not reached in the localized regime.}
\end{figure*}

To get further insights into the dynamical behavior of the system,
we recast the problem in a different language. Starting from the initial
state, denoted as $|0\rangle$, we can generate an orthogonal basis
set $|i\rangle$, $i=0,1,\dots$, by repeatedly applying the Hamiltonian
(see Supplementary Material). In this Lanczos basis of many-body wave
functions, the Hamiltonian has the form of a tight-binding model on
a semi-infinite chain. Each site $i=0,1,\dots$ corresponds to a many-body
state, it has an on-site energy $\epsilon_{i}=\langle i|\mathcal{H}|i\rangle$
and is coupled only to its nearest neighbors by hopping elements $t_{i\to i+1}$
and $t_{i\to i-1}$.~\cite{Lanczos1950} It is easy to realize that
the unitary evolution of the original many body problem is thus fully
equivalent to the dynamics of a single particle, initially sitting
at site $0$, that is then let propagate along such a tight-binding
chain of many-body states. We note that, both $\epsilon_{i}$ and
$t_{i\to i+1}$ largely fluctuate from site to site, therefore resembling
an effective Anderson model, even though those parameters are in reality
deterministic, see Fig.~\ref{fig:1d}. In the same figure, we also
show the mean distance traveled by the particle after time $t$ starting
from the first site of the chain, which corresponds to an initial
state $|0\rangle\equiv(\dots2,0,2,0,\dots)$, and for different $U/J$.
We observe that, for small values of $U/J$, the particle diffuses
and its wave-packet finally spreads over the whole chain, in a rather
uniform way. On the contrary, above a certain critical value of the
interaction $(U/J)_{c}^{dyn}$, the particle stays localized near
the origin for arbitrarily long times. A closer look to the structure
of the on site energies reveal the existence of a potential well at
the edge of the chain. This is crucial in order to understand the
observed localization transition. Since the potential well cannot
induce a true bound state below the bottom of the spectrum, at most
a resonance will form in the spectrum, from which the particle could
in principle escape in a finite time. This indeed happens at small
$U$ but apparently not at large $U$, where the increased depth of
the well and the effective randomness of the on-site energies conspire
together to keep the particle localized close to the edge, preventing
the the states in the well to hybridize with other states along the
chain. We now see how this result connects with the previous analysis
on the density relaxation times. In the small $U/J$ regime the particle
is able to escape from the well and to explore larger portions of
the chain, thus resulting into a fast density relaxation. As opposite,
for large $U/J$, the particle bounces back and forth inside the well,
finding hard time to escape from it. This lack of diffusion results
into a very long-time scale for the density to relax to its homogeneous
value.

The above results show explicitly that some kind of localization in
the many-body configurational space does occur, at least in the finite
system.~\cite{Basko2006,Canovi2010} While such an intriguing behavior
might well be a subtle effect due to the finite size spectrum, it
could also signal the onset of a genuine localization that survives
in the thermodynamic limit.

We conclude the discussion on the inhomogeneous initial states by
studying the case at density $n<1$, when at equilibrium there is
no longer a Mott transition. For example, we consider $n=2/3$ and
an initial density profile $(\dots2,0,0,2,0,0,\dots)$. Interestingly,
we find quite a different behavior for the density relaxation times
$\tau_{R}$ (see Supplementary Material), with a much smoother crossover
from small to large values of $U/J$ and no evidence of any increase
in the relaxation times with the system size. This fact suggests a
deeper connection between the observed dynamical behavior and the
zero temperature Mott transition that occurs at equilibrium and at
integer filling, as suggested by calculations with infinite-coordination
lattices.~\cite{Schiro2010,Sciolla2010}

\subsubsection*{Homogeneous initial states and Quantum Quenches }

\begin{figure}[b]
 \includegraphics[width=0.48\columnwidth]{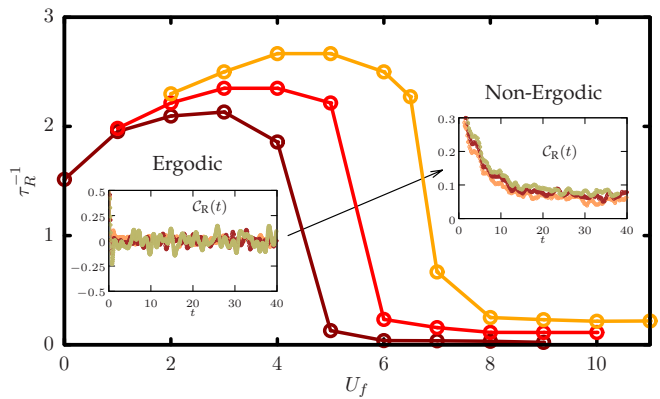} 
\caption{\label{fig:Relax} Inverse relaxation times of density excitations
in the homogeneous system, see Eq.~\eqref{eq:coft}. From left to
right, different curves correspond to different initial states at
$U_{i}/J=0$, $1$, and $2$. Insets: real part of the density correlations
$\mathcal{C}(t)$ in the ergodic and in the non-ergodic region. Data
are obtained with exact-diagonalization on a lattice with $N=12$.}
\end{figure}

In light of the previous results, one may question that by choosing
an inhomogeneous configuration of doublons we pick up a rather specific
initial state in the Hilbert space. We are now going to show that
the above findings strongly affect the dynamics starting from a perfectly
homogeneous state. In this respect, a particularly interesting class
of initial states are the ground states of $\mathcal{H}$ for given
values of the interaction $U_{i}$, which are let evolve under the
Hamiltonian dynamics after a sudden increase of the interaction to
a final value $U_{f}>U_{i}$, the so-called quantum quench. Kollath
and coworkers~\cite{Kollath2007} reported evidence for the existence
of two separated regimes in which either thermal or non-thermal behavior
is observed for local observables. The origin of the non-thermal behavior
in the large $U_{f}$ region and the possibility of an ergodicity
breaking in the thermodynamic limit is still highly debated~\cite{Roux2009,Biroli2010}.

In the following, we focus on an average density $n=1$ and we show
that signatures of long lived metastable states of doblons can be
identified in the dynamics after a quantum quench. At variance with
the previous numerical calculations, now both the initial state and
the quantum Hamiltonian do preserve the spatial homogeneity and, therefore,
the quest for possible signatures of ergodicity breaking requires
a different approach. Since we have identified density relaxation
as the slowest process in the problem, we monitor the dynamics of
the system by measuring the auto-correlation of the density averaged
over all sites, namely through 
\begin{equation}
\mathcal{C}(t)=\frac{1}{N}\sum_{i}\,\left\langle n_{i}(t)n_{i}(0)\right\rangle -\left\langle n_{i}(t)\right\rangle \left\langle n_{i}(0)\right\rangle .\label{eq:coft}
\end{equation}

For any finite size system, $\lim_{t\to\infty}\mathcal{C}(t)=0$ since 
the densities decorrelate at very long times. 
Indeed, as shown in Fig.~\ref{fig:Relax}, for small $U_{f}\ll U_{i}$ this quantity has a 
very fast transient to zero. On the contrary, for $U_{f}\gg U_{i}$ the density
auto-correlation $\mathcal{C}(t)$ gets stuck into a long-lived finite value 
plateau $\mathcal{C}_{\star}\neq0$, before approaching zero only on a much
longer time scale. If we extract a relaxation time from $\mathcal{C}(t)$,
we find a similar behavior as in Fig.~\ref{fig:DecDef}, i.e., a
dramatic increase of the relaxation times above a threshold value
of the final interaction strength.

In agreement with the previous analysis, the appearance of such a
long-lived metastable state characterized by a finite plateau $\mathcal{C}_{\star}$
of the density auto-correlation function might indicate an excess
of double occupancies that have no channel to relax. In other words,
the dynamical constraints brought by the interaction severely slow
down density excitations, whose characteristic time scales increase
abruptly after a critical threshold. The main phenomenological traits
of this \textit{dynamical arrest} characterized by long-lived inhomogeneous
states closely remind the physics of glassy materials.

\subsubsection*{Variational description, lack of thermalization and higher dimensions }

The previous discussion revealed the existence of a threshold energy
above which a steep increase of the relaxation time of density fluctuations
takes place. In order to assess the relevance of this phenomenon for
the dynamics of larger systems or even for higher spatial dimensions,
it is desirable to devise a comprehensive alternative framework able
to catch its very characteristics. Here, we introduce an approach
based on real-time variational Monte Carlo (see Supplementary Material
for details), which has two important advantages: it allows us to
follow the evolution for times comparable to those accessible experimentally,
which are much longer than t-DMRG;~\cite{White2004,Daley2004} it
can be easily extended to higher dimensions. We mention that mean-field-like
variational approaches to the real-time dynamics of correlated systems
have been developed in recent years.~\cite{Schiro2010,Sciolla2010}

\begin{figure}[t]
\includegraphics[width=0.48\paperwidth]{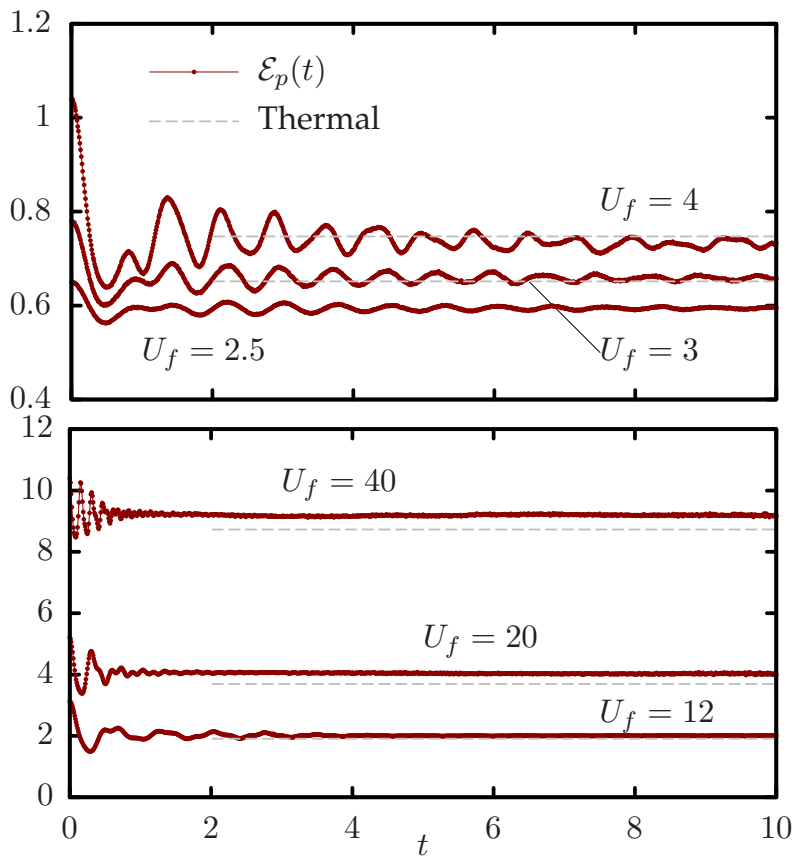} 
\caption{\label{fig:Eq-Values} One-dimensional results for the time-dependent expectation values of the on-site
potential energy $\mathcal{E}_{p}(t)=\frac{U_{f}}{2}\left\langle n_{i}(n_{i}-1)\right\rangle $
in the ergodic (\emph{upper panel}) and in the non-ergodic regions
(\emph{lower panel}). The initial state is the ground state of the
Bose-Hubbard Hamiltonian with $U_{i}=2J$ and the considered system
size is $N=200$. Grand-canonical thermal averages are shown for comparison
as dashed horizontal lines. }
\end{figure}

However, although they seem to capture well the main features of the
dynamical evolution, these methods are unable to describe important
aspects such as damping and relaxation. On the contrary, our approach
is sufficiently rich to account for damping and relaxation of local
observables. It is based on a very simple and transparent out-of-equilibrium
extension of the Jastrow-like variational wave function that was shown
to describe quite accurately the equilibrium phase diagram of the
Bose-Hubbard model:~\cite{Capello2007} 
\begin{equation}
\vert\Psi(t)\rangle=\exp\left(\sum_{ij}\, V_{ij}(n_{i},n_{j};t)\,\right)\,\vert\Psi_{0}\rangle,\label{eq:jastrow}
\end{equation}
 where $\vert\Psi_{0}\rangle$ is the initial state and $V_{ij}(n_{i},n_{j};t)$
is a Jastrow factor that depends on the occupancies $n_{i}$ and $n_{j}$
of two sites $i$ and $j$ and varies with time so to maintain the
time evolution as close as possible to the true evolution via the
Schroedinger equation (see Supplementary Material). The comparison
between our approach and the t-DMRG~\cite{KollathPrivate} is reported
in the Supplementary Materials, demonstrating the high accuracy of
the time-dependent variational Monte Carlo.

Results for the time evolution of a local observable, such as the
potential energy, after a sudden quench from $U_{i}=2J$ to a final
$U_{f}$ are shown in Fig.~\ref{fig:Eq-Values}. The values of the
thermal averages have been computed in the grand-canonical ensemble
by means of finite-temperature quantum Monte Carlo calculations,~\cite{Alps2011}
with the effective temperature fixed by the average energy of the
initial state, which we take as the best variational approximation
for the ground state at $U_{i}=2J$.

As shown in Fig.~\ref{fig:Eq-Values}, in the region of small $U_{f}$
we observe a damping of the average potential-energy, which approaches
a quasi-steady stationary value in contrast to the simple Gutzwiller
wave function,~\cite{Schiro2010,Sciolla2010}. In this regime, damping
is mainly due to a density-density Jastrow factor of the form 
$V_{ij}(n_{i},n_{j};)=v_{ij}(t) \, n_{i}\, n_{j}$,
which already at equilibrium was shown to provide a satisfactory description
of the physical behavior.~\cite{Capello2007} This fact enlightens
the relevance of the Bogoliubov modes whose dephasing during the time
evolution allows to approach the stationary state. Remarkably, the
steady state averages coincide with the thermal ones; a signal that
the dynamics is ergodic.

In the region of large interactions $U_{f}$, a simple density-density
Jastrow factor does not account for all relaxation pathways, which
will now mainly result from specific correlations among doublons, holons 
and between holons and doublons. The effective Hamiltonian~\eqref{eq:Heff}
indeed explicitly shows that doublons attract each other as well as
holons do, while doublons repel holons. These correlations, as well
as other among higher on-site densities, can be easily implemented
via the the Jastrow factor in Eq.~\eqref{eq:jastrow} and indeed
substantially improve the dynamics. Interestingly, the effective interaction
between doblons that results from the dynamical variational calculation
turns to be \emph{attractive}, therefore leading to a consistent determination
of the anticipated dynamical effects that drive the dynamics in this
regime. As we see from Fig.~\ref{fig:Eq-Values}, in the region of
very large $U_{f}$ the potential-energy expectation values do show
a damping to a non-thermal quasi-steady state on a time of the order
of $\tau_{D}\sim1/U_{f}$. This fast time scale must be put in comparison
with the much longer one, $\tau_{R}$ of Fig.~\ref{fig:Relax}. It
is natural to identify $\tau_{R}$ with the time scale that controls
the eventual escape from the quasi-steady state, hence the approach
to thermal equilibrium. Whether this time scale does truly diverge
in the thermodynamic limit, or rather saturates to a very large value
which is still out of reach for state of the art numerics, it is certainly
an important issue that cannot be definitively solved. However, we
can safely state that a large finite system of actual experimental
relevance will get stuck for a quite long time into highly inhomogeneous
metastable states, which we revealed to be on the verge of a spatial
symmetry breaking.

\begin{figure}
\includegraphics[width=0.45\paperwidth]{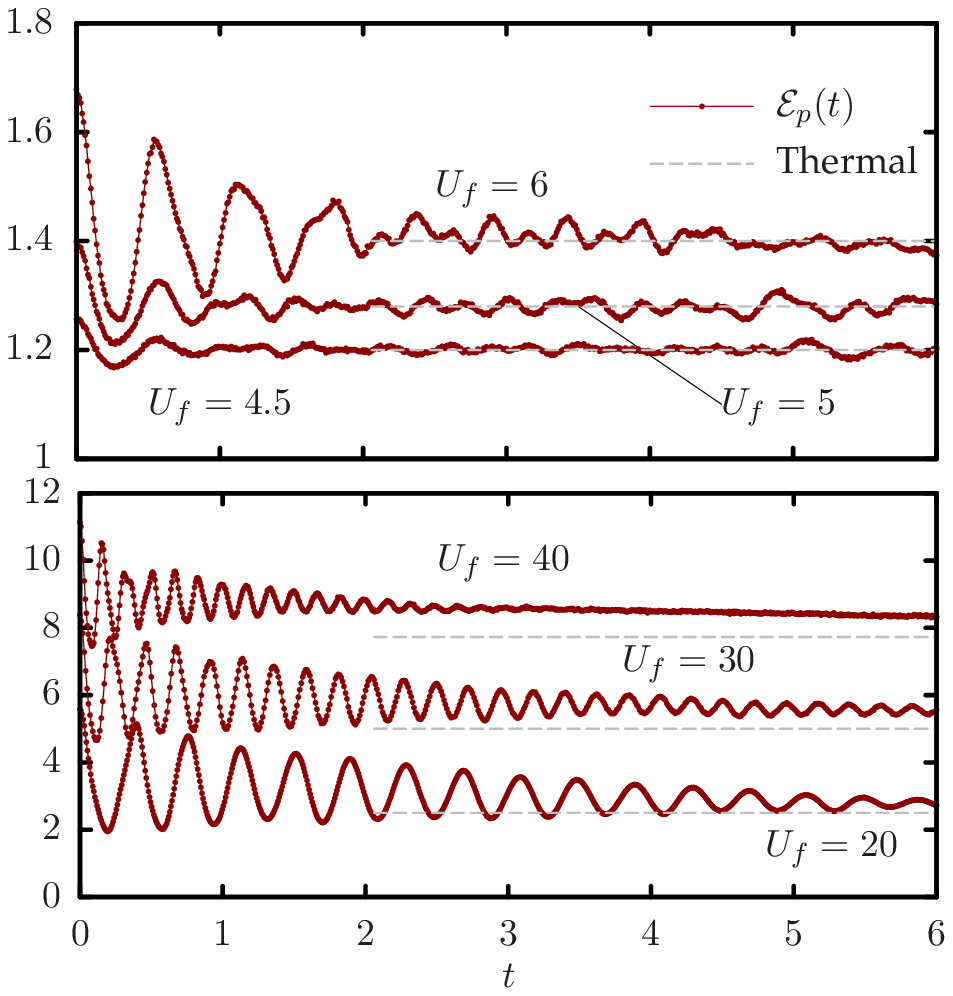} 
\caption{\label{fig:Eq-Values-2d} Two dimensional results for the time-dependent expectation values of the on-site
potential energy $\mathcal{E}_{p}(t)=\frac{U_{f}}{2}\left\langle n_{i}(n_{i}-1)\right\rangle $
in the ergodic (\emph{upper panel}) and in the non-ergodic regions
(\emph{lower panel}). The initial state is the ground state of the
Bose-Hubbard Hamiltonian with $U_{i}=4J$ and the considered system
size is $N=20\times20$. Grand-canonical thermal averages are shown for comparison
as dashed horizontal lines. }
\end{figure}

The above results have been obtained for out-of-equilibrium one dimensional
systems and therefore leave the questions concerning the dependence
on the dimensionality of the problem still open. However, the anomalously
long-time relaxation of the density auto-correlation points towards
a kind of glassy behavior that should be observable even in higher
dimensions, provided that the interaction induces sufficiently strong
dynamical constraints. In this regard, we have studied the two-dimensional
case by means of our time-dependent variational scheme, verifying that a 
similar behavior occurs even in two dimensions. In Fig.~\ref{fig:Eq-Values-2d},
we show the results of the potential-energy expectation values as a function
of the time. As before, if the repulsion is weak, the time average coincides
with the thermal ones, while for strong repulsion there is a clear difference
between thermal and time averages, giving rise to a nonergodic dynamics.
We therefore argue that the phenomenology we have hereby identified is almost 
independent on the dimensionality and it is rather due to the existence in 
strongly correlated systems of high-energy incoherent excitations that do not
have channels to relax efficiently. This scenario is also consistent
with the observed strong doping dependence of the relaxation time
(see Supplementary Material).

\section*{Discussion}

In conclusion, we have found that the dynamical constraints brought
by a strong interaction can trap the evolution of repulsive bosons
hopping on a lattice into metastable states that lack translational
symmetry, provided that the energy stored into the initial state is
above a threshold. We pointed out that a self-induced effective attraction
among doublons is one of the major processes that can effectively
freeze the dynamics on long time scales. Such a mechanism is recognized
to play a role in the density relaxation processes of purely homogeneous
systems through a dynamical arrest visible in time-dependent density
correlations. The main features of this intriguing behavior, namely
the slowing down of density excitations and the long-lived inhomogeneous
pattern, resembles closely a kind of glass transition.

Moreover, we have shown that the time evolution of the many-body problem
can be mapped onto that of a particle moving from the edge of a semi-infinite
tight-binding chain with nearest-neighbor hopping, where each site
represents a many-body wave function. This model looks like an Anderson
model, since both the on-site energy and the hopping vary from site
to site, with a potential well at one edge due to the high-energy
content of the initial state. Interestingly, we find a delocalization-localization
transition in this problem, with the particle being unable to diffuse
on the whole chain above a certain value of the well depth. We consider
this analogy quite suggestive and potentially constitute an even stronger
indication of ergodicity breaking in the many-body space which is
worth to be further investigated.

\subsubsection*{Acknowledgments}

We acknowledge discussions with G. Biroli, J.F. Carrasquilla, D. Huse,
A. Parola, E. Tosatti, F. Zamponi, and C. Kollath also for providing
us with t-DMRG numerical data. Computational support from CASPUR through
the standard HPC Grant 2011 is also acknowledged.

\subsubsection*{Author contributions}

All authors conceived and designed the research, equally contributing
to the preparation of the manuscript. G.C. also designed and carried
out the numerical calculations.

\subsubsection*{Competing Financial Interests}

The authors declare no competing financial interests.

\newpage{}

\section*{\emph{--Supplementary Material--}\protect \\
}

\subsection*{Exact Diagonalization }

The exact time evolution of an initial quantum state $|\Psi(0)\rangle$
is given by: 
\begin{equation}
|\Psi(t)\rangle=e^{-i\mathcal{H}\; t}|\Psi(0)\rangle=\sum_{n}c_{n}e^{-iE_{n}\; t}|\Phi_{n}\rangle,
\end{equation}
 where $|\Phi_{n}\rangle$ are the eigenstates of $\mathcal{H}$,
with corresponding eigenvalues $E_{n}$, and $c_{n}=\langle\Phi_{n}|\Psi(0)\rangle$.
Therefore, in order to follow the quantum dynamics to large times,
the full knowledge of the spectrum is in principle needed. However,
whenever the Hilbert space is so large that it does not allow for
a full diagonalization, an alternative approach is in order. At each
time, we can make an evolution for a time step $\Delta t$ by considering
a truncated Taylor expansion of the unitary evolution: 
\begin{equation}
|\Psi(t+\Delta t)\rangle=e^{-i\mathcal{H}\;\Delta t}|\Psi(t)\rangle\simeq\sum_{k=0,k_{max}}\frac{(-i\Delta t)^{k}}{k!}\mathcal{H}^{k}|\Psi(t)\rangle.
\end{equation}
 Here, the series converges very quickly with $k$ and the cutoff
$k_{max}$ must be chosen to obtain the desirable convergence for
a given $\Delta t$. Therefore, the evolved wave function $|\Psi(t+\Delta t)\rangle$
can be easily obtained by summing terms that can be in turn recovered
by repeatedly applying the Hamiltonian $\mathcal{H}$ to $|\Psi(t)\rangle$.
The full time evolution at long times can be achieved by subsequent
small-time evolutions.

Moreover, time-dependent correlation functions of the form $\mathcal{C}_{A}(t)=\langle\Psi|A^{\dagger}(t)A(0)|\Psi\rangle$,
where $\mathcal{A}$ is an arbitrary operator, can be as well obtained
by first applying $A$ to $|\Psi\rangle$, then performing the time
evolution $|\Psi_{A}(t)\rangle=e^{-i\mathcal{H}\; t}A|\Psi\rangle$,
and finally computing $\mathcal{C}_{A}(t)=\langle\Psi(t)|A^{\dagger}|\Psi_{A}(t)\rangle$,
where $|\Psi(t)\rangle=e^{-i\mathcal{H}\; t}|\Psi\rangle$.

The numerical accuracy of the unitary evolution can be verified by
checking the conserved quantities of the time-evolution (for example
the total energy) which in our calculations remain to all purposes
constant up to the longest considered evolution times.

\subsection*{Lanczos Method}

In this work we have used the Lanczos method to construct a basis
of $L$ orthonormal states in which the full Hamiltonian takes a reduced
tridiagonal form: 
\begin{equation}
\mathcal{H}_{L}=\begin{pmatrix}\epsilon_{0} & t_{0\to1} & 0 & \vdots & 0\\
t_{1\to0} & \epsilon_{1} & t_{1\to2} & \vdots & 0\\
0 & t_{2\to1} & \epsilon_{2} & \vdots & 0\\
0 & 0 & \dots & \ddots & t_{L-1\to L}\\
0 & 0 & \dots & t_{L\to L-1} & \epsilon_{L}
\end{pmatrix}\label{eq:tridiag}
\end{equation}
 The basis set is constructed in a recursive way starting from an
initial vector $\left|0\right\rangle $ upon repeatedly applying the
Hamiltonian 
\begin{eqnarray}
t_{n\to n+1}\left|n+1\right\rangle  & = & \mathcal{H}\left|n\right\rangle -\epsilon_{n}\left|n\right\rangle -t_{n-1\to n}\left|n-1\right\rangle ,\label{eq:lancstates}
\end{eqnarray}
 with the coefficients of the tridiagonal matrix given by 
\begin{eqnarray}
t_{n\rightarrow n+1} & = & \left\langle n+1\right|\mathcal{H}\left|n\right\rangle \label{eq:lanccoeff}\\
\epsilon_{n} & = & \left\langle n\right|\mathcal{H}\left|n\right\rangle .\nonumber 
\end{eqnarray}
 Once the tridiagonal matrix is constructed we are in condition to
consider a time-dependent problem of a particle initially localized
on site $\left|0\right\rangle $ whose dynamics is governed by $\mathcal{H}_{L}$.

A final important remark concerns the practical numerical evaluation
of the Lanczos states and matrix elements on calculators with a finite-precision
arithmetic. Numerical truncation errors can indeed accumulate and
lead the most recently generated Lanczos vector not to be orthogonal
to the previously generated ones. To overcome this problem, we have
carefully analyzed the dependence of the results on the machine precision,
considering up to $10^{4}$ bits of floating point arithmetic, for
which no significant truncation error manifests up to the largest
value of $L$ considered in the paper.

\subsection*{Time-Evolving Block decimation}

The general idea of the Time-Evolving Block decimation algorithm~\cite{Vidal2004,TebdOpen}
is to consider a particular representation of a generic many-body
state defined in a Hilbert space of dimension $\mathcal{D}^{N}$,
where $\mathcal{D}$ is the dimension of a certain local-basis. Such
a generic state is fully described in terms of the local quantum numbers
$i_{k}$ as

\begin{eqnarray*}
|\Psi\rangle & = & \sum\limits _{i=1}^{\mathcal{D}}c_{i_{1}i_{2}..i_{N}}|{i_{1},i_{2},..,i_{N-1},i_{N}}\rangle,
\end{eqnarray*}
 and in the Vidal's representation the coefficients $c_{i_{1}i_{2}..i_{N}}$
are taken to be 
\begin{eqnarray}
c_{i_{1}i_{2}..i_{N}} & = & \sum\limits _{\alpha_{0},..,\alpha_{N}}^{\chi}\lambda_{\alpha_{0}}^{[1]}\Gamma_{\alpha_{0}\alpha_{1}}^{[1]i_{1}}\lambda_{\alpha_{1}}^{[2]}\Gamma_{\alpha_{1}\alpha_{2}}^{[2]i_{2}}\lambda_{\alpha_{2}}^{[3]}\Gamma_{\alpha_{2}\alpha_{3}}^{[3]i_{3}}\lambda_{\alpha_{3}}^{[4]}\cdot..\cdot\lambda_{\alpha_{N-1}}^{[N]}\Gamma_{\alpha_{N-1}\alpha_{N}}^{[N]i_{N}}\lambda_{\alpha_{N}}^{[N+1]},\label{eq:cvidal}
\end{eqnarray}
 where the local tensors $\Gamma$ and the \emph{bond} matrices $\lambda$
constitute a set of $\mathcal{D}\chi^{2}N+\chi\left(N+1\right)$ parameters
that specify the state in this representation. The time evolution
of a given initial state is obtained upon considering a small time-step
$\Delta t$, and a repeated application of the real-time propagator
$e^{-i\mathcal{H}\Delta t}$ within a Suzuki-Trotter approximation
scheme, inducing systematic errors of magnitude $\mathcal{O}\left(\Delta t^{3}\right)$.
When the infinitesimal unitary evolution is applied the decomposition~\eqref{eq:cvidal}
is consistently updated. The other major source of systematic errors
is however intrinsically due to the finite amount of entanglement
retained by Vidal's representation. We have checked the convergence
of the studied properties with respect to both the time step and the
entanglement cutoff parameter, considering a maximum of $\chi=400$,
for which the density is converged up to the plotted times of Fig.
2 (right panel) in the main paper.

\subsection*{Real-time variational Monte Carlo}

In this part of the Supplementary Materials, we give the details of
the real-time variational Monte Carlo approach. First of all, we would
like to stress that this method does not suffer from intrinsic instabilities
arising from the sign (or phase) problem. Indeed, our method is essentially
a straightforward generalization of the variational technique, where
the probability density is strictly non-negative. Although approximated,
this approach has firm physical basis and can accurately reproduce
numerically exact results in cases where t-DMRG method can be applied
(e.g., mostly in one-dimensional lattices). Nonetheless, variational
Monte Carlo can be generalized and applied in larger dimensions without
any further numerical instability.

\begin{figure}
\includegraphics[width=0.5\columnwidth]{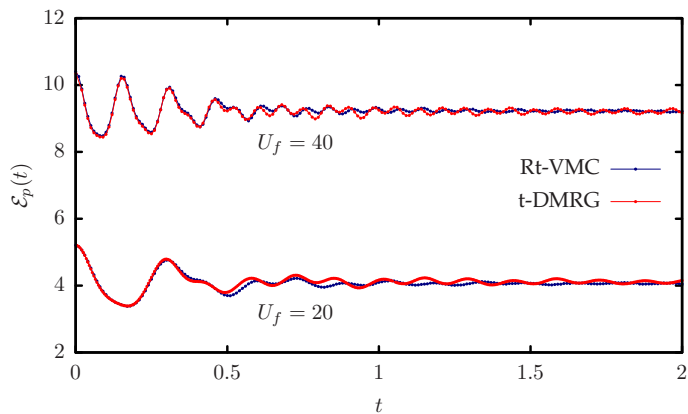} \caption{\label{fig:comparison} Time-dependent expectation values of the on-site
potential energy $\mathcal{E}_{p}(t)=\frac{U_{f}}{2}\left\langle n_{i}(n_{i}-1)\right\rangle $.
The initial state is the ground state of the Bose-Hubbard Hamiltonian
with $U_{i}=2J$. Real-time Variational Monte Carlo results are obtained
for an $N=100$ chain with periodic boundaries. Corresponding data
by t-DMRG \cite{KollathPrivate} are the potential energy at the center
of an $N=64$ chain with open boundaries. }
\end{figure}

In general, it is reasonable to expect that the relevant physical
properties driving the dynamics can be reduced to a number of parameters
significantly smaller than the size of the full Hilbert space of the
problem, as much as it is done in renormalization-based numerical
and analytical tools. Let us introduce a set of {}``excitation operators''
$\mathcal{O}_{k}$, which are expected to describe the relevant excitations
over the ground state that are involved in the dynamics of the system.
The actual set of these operators has to be carefully chosen, depending
upon the properties of the quantum phase. In the case of one-dimensional
interacting bosons, $\mathcal{O}_{k}$ can be selected to be density-density
correlations, e.g., $\mathcal{O}_{k}=\sum_{i}n_{i}n_{i+k}$, or doublon-doublon,
holon-holon, and doublon-holon correlations. As discussed in the text,
the presence of the former ones is sufficient to obtain a very accurate
time dynamics for small quenches, whereas the inclusion of the latter
ones is necessary for large quenches. In Fig.~\ref{fig:comparison},
we report the comparison of the t-DMRG dynamics with our variational
Monte Carlo approach.

Let us discuss the details of our method. First of all, we define
a basis $|\mathbf{x}\rangle$ in which the occupation numbers of the
bosons are defined in the lattice. Since the operators $\mathcal{O}_{k}$
are diagonal in this basis, we will define $\mathcal{O}_{k}(\mathbf{x})$
as the (time-independent) value that the operator takes over the configuration
$|\mathbf{x}\rangle$. The variational wave function is given by:
\begin{equation}
\Psi(\mathbf{x},t)=e^{\theta(t)}\exp\left[\sum_{k}^{N_{p}}\mathcal{O}_{k}(\mathbf{x})\alpha_{k}(t)\right]\xi(\mathbf{x}),\label{eq:psivar}
\end{equation}
 where $\xi(x)$ is a time-independent uncorrelated state, such as
the bosonic condensate, whereas $\theta(t)$ and $\alpha_{k}(t)$
are time-dependent complex parameters.

The variational form introduced here allows us to map the quantum
dynamics associated to a given Hamiltonian $\mathcal{H}$ into the
dynamics of the variational parameters $\theta(t)$ and $\alpha_{k}(t)$.
The operators $\mathcal{O}_{k}$ can be thought as collective variables
for the quantum dynamics; on physical grounds, we expect that the
quantum dynamics will be characterized by a set of collective variables
whose number $N_{p}$ is much smaller than the original size of the
Hilbert space. This reduction assumption is the key requirement for
having an efficient time-dependent variational approach.

The equations of motion for the variational parameters can be found
by minimizing the Euclidean distance in the Hilbert space between
the \textit{exact} and the variational time dynamics. Given at a certain
time the quantum state $\Psi(\mathbf{x},t)$, its exact time evolution
leads to $\dot{\phi}(\mathbf{x},t)=\Psi(\mathbf{x},t)\times\left[-i\mathcal{E}(\mathbf{x},t)\right]$,
where the complex-valued {}``local-energy'' is $\mathcal{E}(\mathbf{x},t)=\frac{\langle\mathbf{x}|\mathcal{H}|\Psi(t)\rangle}{\Psi(\mathbf{x},t)}$.
On the other hand, an infinitesimal change in time in the variational
parameters leads to $\dot{\Psi}(\mathbf{x},t)=\Psi(\mathbf{x},t)\times\left[\dot{\theta}(t)+\sum_{k}\mathcal{O}_{k}(\mathbf{x})\dot{\alpha}_{k}(t)\right]$.
The minimization of their mutual distance $\mathcal{D}(t)=\sum_{\mathbf{x}}\left|\dot{\phi}(\mathbf{x},t)-\dot{\Psi}(\mathbf{x},t)\right|^{2}$
gives a set of coupled differential equations for $\dot{\alpha_{k}}(t)$
and $\dot{\theta}(t)$: 
\begin{eqnarray}
 &  & \sum_{k'}\langle\delta\mathcal{O}_{k'}\delta\mathcal{O}_{k}\rangle\dot{\alpha}_{k'}(t)=-i\langle\mathcal{E}(t)\delta\mathcal{O}_{k}\rangle\label{eq:eqmotion1}\\
 &  & \dot{\theta}(t)=-i\langle\mathcal{E}(t)\rangle-\sum_{k'}\dot{\alpha}_{k'}(t)\langle\mathcal{O}_{k'}\rangle,\label{eq:eqmotion2}
\end{eqnarray}
 where $\delta\mathcal{O}_{k}=\mathcal{O}_{k}-\langle\mathcal{O}_{k}\rangle$
and $\langle\dots\rangle$ denote expectation values over the square
modulus of the wave function at time $t$, namely $\langle F\rangle=\frac{\sum_{\mathbf{x}}\left|\psi(\mathbf{x},t)\right|^{2}F(\mathbf{x})}{\sum_{\mathbf{x}}\left|\psi(\mathbf{x},t)\right|^{2}}$.

The coupled equations of motion determine a consistent Hamiltonian
dynamics in which at all times both the norm $\mathcal{N}(t)=\sum_{\mathbf{x}}\left|\psi(\mathbf{x},t)\right|^{2}$
and the expectation value of the energy $E(t)=\langle\mathcal{E}(t)\rangle$
are strictly conserved. Indeed, one can show that $\dot{N}(t)\propto\left[\dot{\theta}^{R}+\sum_{k}\dot{\alpha}_{k}^{R}(t)\langle\mathcal{O}_{k}\rangle\right]$
and $\dot{E}(t)\propto\sum_{k}\left\{ \dot{\alpha}_{k}^{R}\langle\delta\mathcal{O}_{k}\mathcal{E}^{R}\rangle+\dot{\alpha}_{k}^{I}\langle\delta\mathcal{O}_{k}\mathcal{E}^{I}\rangle\right\} $,
which are both vanishing for the optimal solutions for which $\mathcal{D}(t)$
is at a minimum.

Not surprisingly, the same equation of motions that we have derived
by means of an optimal distance principle can be also derived from
the variational principle, namely it can been shown that the action
$\mathcal{S}=\int dt\langle\Psi\left|i\frac{\partial}{\partial t}-\mathcal{H}\right|\Psi\rangle$
is indeed stationary for the time-trajectories of the variational
parameters induced by Eqs.~\eqref{eq:eqmotion1} and~\eqref{eq:eqmotion2}.

The time-dependent expectation values over the square modulus of the
many-body wave function cannot be calculated analytically for generic
operators $\mathcal{O}_{k}$, and a variational quantum Monte Carlo
is needed to evaluate them. In particular, $\left|\psi(\mathbf{x},t)\right|^{2}$
can be straightforwardly interpreted as a probability distribution
over the Hilbert space spanned by the configurations $\mathbf{x}$
and a Markov process can be devised, such that the stationary equilibrium
distribution coincides with the desired probability measure.

\subsection*{Inhomogeneous states dynamics at $\mathbf{n=2/3}$ }

\begin{figure}
\includegraphics[width=0.5\columnwidth]{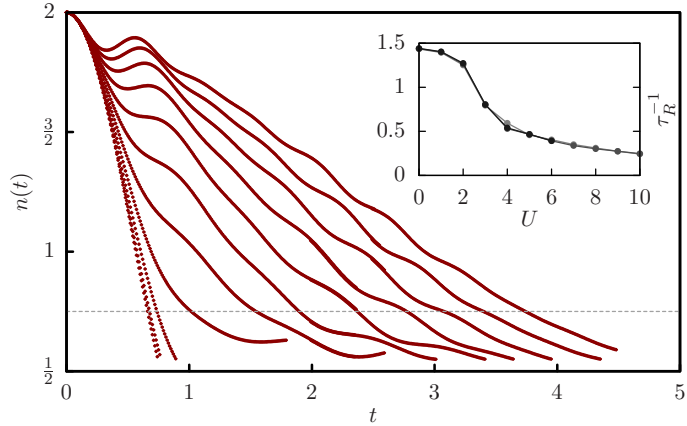} \caption{\label{fig:neq1} Exact time evolution of the densities on initially
doubly occupied sites at filling $n=2/3$ for an $N=15$ chain. Different
curves from left to right correspond to increasing values of the final
interaction at $U/J=0,1,\dots10$. Inset: corresponding inverse relaxation
times $\tau_{R}^{-1}$ for increasing lattice size $N=9$, $12$,
and $15$.}
\end{figure}

In the main paper, we have concentrated our attention on the unitary-filling
density, where at equilibrium a Mott transition takes place upon increasing
the interaction strength. To better realize the importance of the
density on our considerations on the large-$U$ relaxation times,
we now show the exact time evolution of inhomogeneous states at $n<1$.
For example, we consider $n=2/3$ and an initial density profile $(\dots2,0,0,2,0,0,\dots)$.
As shown in Fig.~\ref{fig:neq1}, we find quite a different behavior
for the density relaxation times with respect to the unitary-filling
case, with a much smoother crossover from small to large values of
$U_{f}$. Moreover, there is no evidence of any increase in the relaxation
times with the system size. This fact suggests that the dynamical
constraints brought by the effective interaction among doblons are
much stronger at $n=1$, where a sharp crossover in the relaxation
times is observed. We finally remark that, even at non unitary fillings,
we expect larger clusters of doblons to have larger relaxation times
than the simple initial state we have considered here. The possibility
of a non-thermal behavior at non-unitary filling cannot be excluded,
even though the strongest manifestation of the dynamical arrest is
expected to happen at $n=1$.

\subsection*{Lanczos-basis analysis of an integrable model}

\begin{figure*}
\includegraphics[width=0.42\columnwidth]{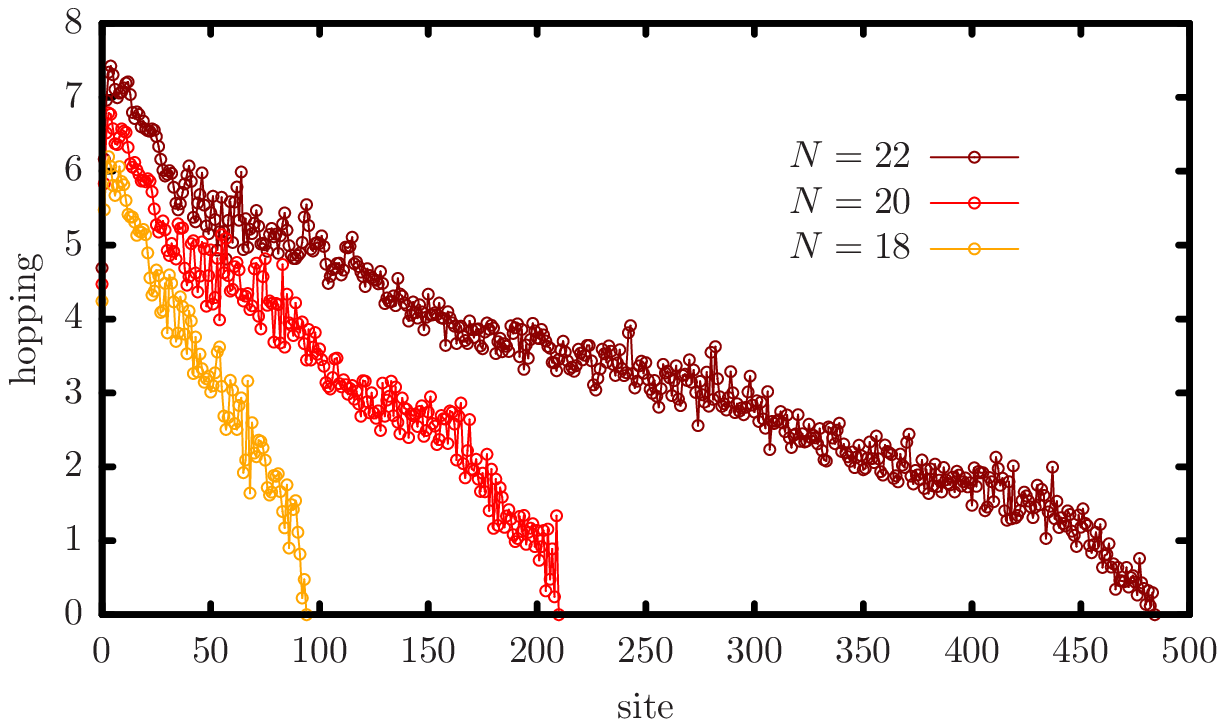}\hfill{}\includegraphics[width=0.42\columnwidth]{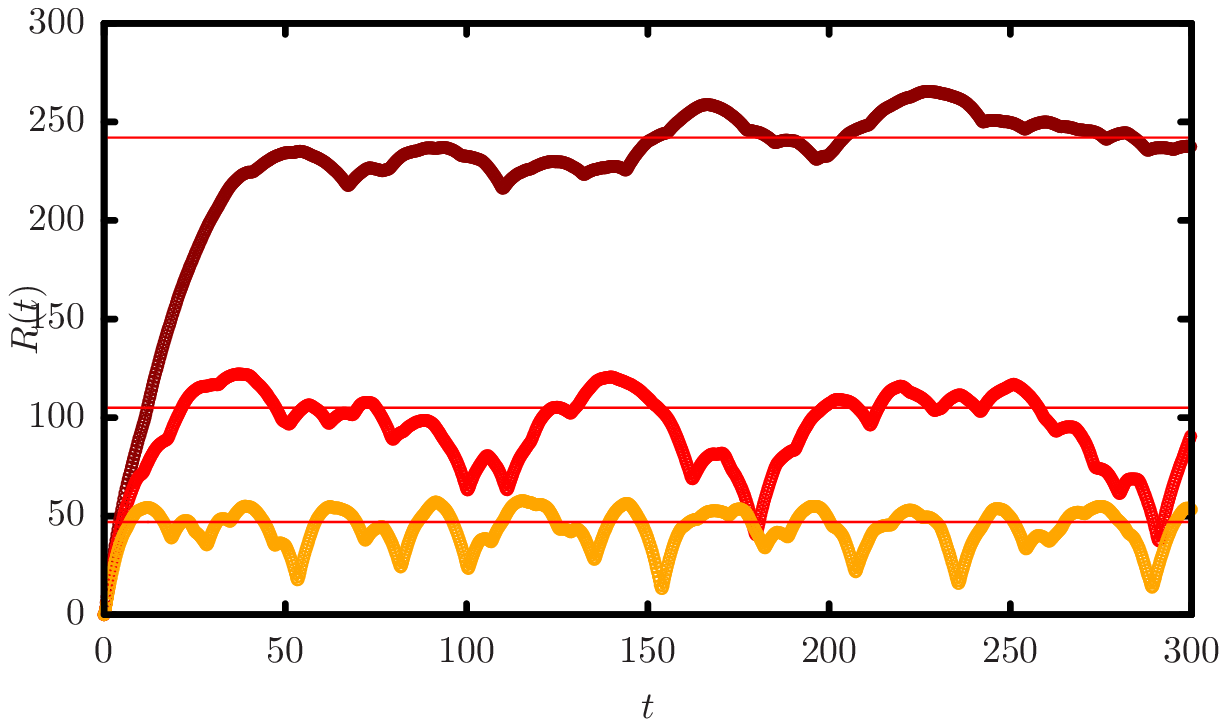}
\caption{\label{fig:V0mod} \emph{Left panel} -- Nearest neighbor hopping of
the effective chain that represents the non-interacting hard-core
bosons Hamiltonian in the Lanczos basis starting from $(\dots1,0,1,0,\dots)$
\emph{Right panel} -- Time-dependent expectation value of the wave-packet
position traveling in the Hilbert space generated by a chain of $\nu(N)$
Lanczos states. The horizontal lines correspond to $\nu(N)/2$ and
show that delocalization takes place.}
\end{figure*}

To better elucidate the many-body localization phenomenon suggested
by the Lanczos-basis analysis of the Bose-Hubbard model, we hereby
show that localization is \emph{not} strictly implied by integrability.
To this purpose, we consider a model of one-dimensional hard-core
bosons prepared in an initial inhomogeneous state with an alternating
density profile $(\dots1,0,1,0,\dots)$ and let it evolve in the Lanczos
basis with their non-interacting (integrable) kinetic Hamiltonian.
In Fig.~\ref{fig:V0mod}, we show both the Lanczos hopping elements
and the expectation value of the Lanczos particle traveling in the
many-body space. The effect of integrability is to reduce the number
of allowed Lanczos states with respect to the total Hilbert space.
Indeed, we find that the hopping is a strongly decreasing function
of the iterations, approaching zero after a certain number of states
$\nu(N)$ that is a finite fraction of the full Hilbert space. Noticeably,
the average position traveled by the particle increases accordingly,
leading, therefore, to a full delocalization in the thermodynamic
limit. This analysis shows that a localization in the Lanczos basis
is not necessarily due to integrability -- whose effect only amounts
to reduce the number of active many-body states -- whereas it is due
to the effective dynamical constraints brought by the interaction.

\end{document}